\documentclass[aps,prb, twocolumn, nolongbibliography, nofootinbib, superscriptaddress,10pt, noeprint]{revtex4-2}

\makeatletter
\g@addto@macro\bfseries{\boldmath}
\makeatother

\usepackage{amsmath,amssymb,graphicx,dcolumn,bm,hyperref,float}
\usepackage{microtype}
\hypersetup{ colorlinks=true, linkcolor=red, filecolor=red, urlcolor=red, citecolor = red}
\usepackage{xcolor}
\usepackage{csquotes}
\usepackage{tikz}
\usepackage{amsthm, mathtools}
\usepackage[caption=false]{subfig}
\usepackage{braket}
\usepackage[normalem]{ulem}
\usepackage{setspace}

\newcommand{\beq}{\begin{equation}}
\newcommand{\eeq}{\end{equation}}
\newcommand{\qe}{\varepsilon}
\newcommand{\df}{d_f} 
\newcommand{\rmax}{r_{\it max}}
\newcommand{\emin}{\qe_{\it min}}
\newcommand{\prob}{\mathbb P}
\newcommand{\expect}{\mathbb E}
\newcommand{\PPP}{P_{\it{PP}}}
\newcommand{\subhead}[1]{\textit{#1.---}}
\newcommand{\HH}{{\bf H}}
\newcommand{\boldups}{{\bf \Upsilon}}
\newcommand{\GG}{\mathcal{G}}
\newcommand{\VV}{\mathcal{V}}
\newcommand{\EE}{\mathcal{E}}
\newcommand{\MM}{\mathcal{M}}

\DeclareMathOperator*{\Pf}{Pf}
\DeclareMathOperator*{\sgn}{sgn}
\DeclareMathOperator*{\spec}{spec}
\DeclareMathOperator*{\argmin}{argmin}

\graphicspath{{./figures/}}

\begin{document}

\title{Low-temperature transition of 2d random-bond Ising model\\ and quantum infinite randomness }

\author{Akshat Pandey}
\thanks{These authors contributed equally to this work.}
\affiliation{Department of Physics, Stanford University, Stanford, CA 94305, USA}
\affiliation{All Souls College, Oxford, OX1 4AL, United Kingdom}
\affiliation{Rudolf Peierls Centre for Theoretical Physics, University of Oxford, Oxford, OX1 3PU, United Kingdom}

\author{Aditya Mahadevan}
\thanks{These authors contributed equally to this work.}
\affiliation{Department of Physics, Stanford University, Stanford, CA 94305, USA}
\affiliation{National Institute for Theory and Mathematics in Biology, Chicago, IL 60611, USA}

\author{A.~Alan Middleton}
\affiliation{Department of Physics, Syracuse University, Syracuse, New York 13244, USA}

\author{Daniel S.~Fisher}
\affiliation{Department of Applied Physics, Stanford University, Stanford, CA 94305, USA}

\date{\today}

\begin{abstract}
At low temperatures, the classical two-dimensional random bond Ising model undergoes a frustration-driven ferromagnet-to-paramagnet transition controlled by a zero-temperature fixed point separating ferromagnet and spin glass phases. We show that this critical point can be understood through a renormalization group transformation that constructs the ground state of the Ising model through a sequence of Hamiltonians that, starting with an unfrustrated model, iteratively adds in frustration until the target Hamiltonian is reached. Via a mapping of the thermodynamics of the 2d Ising model to the spectral properties of a related Hermitian matrix---the Hamiltonian of a noninteracting quantum problem---this RG procedure corresponds to an iterative diagonalization of the quantum Hamiltonian. The flow toward zero temperature in the Ising picture manifests as a flow toward infinite randomness in the spectrum of the quantum Hamiltonian, with the log gap of the Hamiltonian scaling as a power of the system size: $\log \emin^{-1} \sim L^\psi$. The tunneling exponent $\psi$ is equal to the spin stiffness exponent $\theta_c$ characterizing the zero-temperature fixed point. 
\end{abstract}

\maketitle

\begin{figure}
 \centering
 \includegraphics[width=0.8\columnwidth]{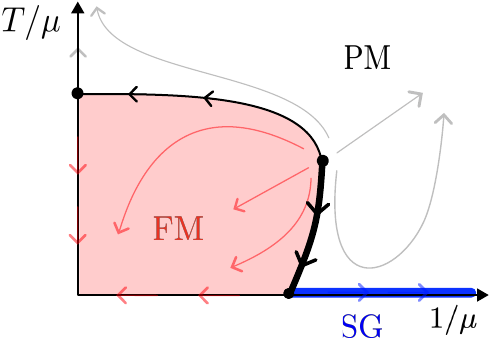}
 \caption{
 Qualitative phase diagram of the 2d random-bond Ising model, tuned by temperature $T$ and the inverse of the mean $\mu$ of the distribution of couplings, showing ferromagnet (FM), paramagnet (PM), and $T=0$ spin glass (SG) phases. Arrows are RG flows, with colors corresponding to destination phases. Black arrows are critical flows. The thick black line indicating the low-$T$ FM-SG transition below the multicritical point, and particularly the $T=0$ fixed point which controls it, are our focus here.
 }
 \label{fig:phase_diagram}
\end{figure}

\subhead{Introduction} 
Classical Ising models on planar graphs (in particular a square lattice), with Hamiltonian $\mathcal{H} = -\sum_{\langle ij \rangle}J_{ij} \sigma_i \sigma_j$ for bonds $\langle ij\rangle$ and spins $\sigma_i \in \{-1,1\}$, can be studied through mappings to various ``spaces''~\cite{Onsager, SchultzMattisLieb, McCoyWuBook, BaxterBook}. 
Of particular interest---still only partially understood---are models with random couplings drawn from various forms of joint distributions. 

If the randomness is only in one direction with the $J_{ij}$ constant in the other direction~\cite{McCoyWuThermodynamics}, a well-known mapping of the non-random direction to imaginary time relates the 2d Ising transfer matrix to a random quantum transverse-field Ising chain~\cite{SchultzMattisLieb}. 
At the critical point of this model, an asymptotically exact renormalization group (RG) in the quantum representation yields a distribution of effective couplings at low energies that broadens without bound on a logarithmic scale---the critical RG fixed point lies at ``infinite randomness''~\cite{FisherTFIM}. This leads, in contrast to the conventional scaling $\emin \sim L^{-z}$ of the gap $\emin$ with system size $L$, to ``tunneling'' dynamical scaling $\log \emin^{-1} \sim L^\psi$---a particular form of $z=\infty$---with $\psi$ an RG eigenvalue that measures the rate of flow to infinite randomness.

\begin{figure}
 \centering
 \includegraphics[width=\columnwidth]{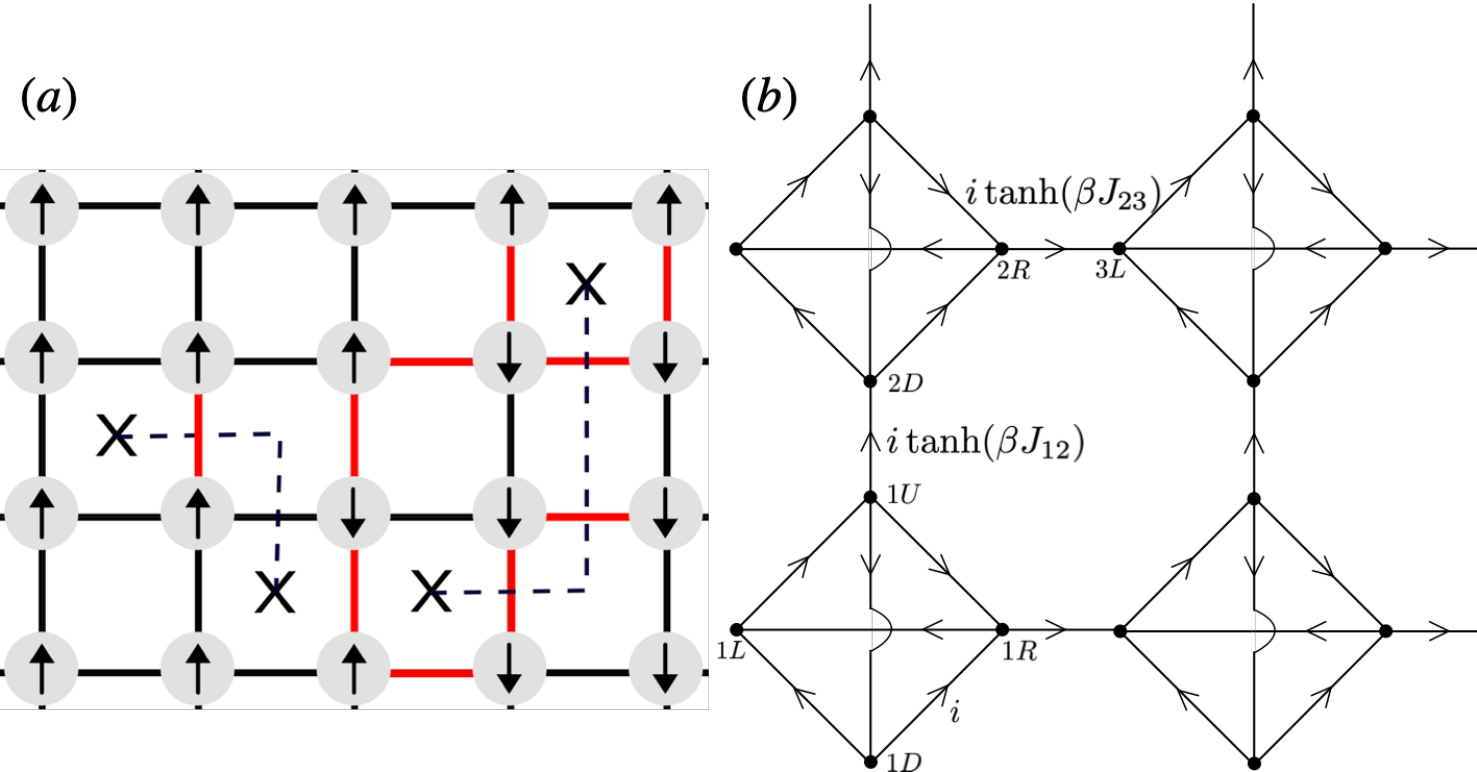}
 \caption{Classical and quantum representations. (a) A spin configuration, and the corresponding set of unsatisfied strings (dashed) on the dual lattice that cross unsatisfied bonds and terminate on frustrated plaquettes (crosses). Antiferromagnetic bonds are shown in red. 
 (b) Four unit cells, corresponding to four Ising spins, of the lattice on which the antisymmetric pure-imaginary $\HH$ lives. 
 The four-site ``city" $1\{R,U,D,L\}$ surrounds spin $1$, etc. 
 Elements of $\HH$ are shown, with orientation dictating sign: for example, $H_{1D,1R}=-H_{1R,1D}=i$ (all intra-city elements are $i$ in the direction of arrows), and $H_{1U,2D}=-H_{2D,1U}=i\tanh(\beta J_{12})$.  }
 \label{fig:definitions}
\end{figure}

For a {\it classical} 2d Ising model, however, it is much more natural to consider i.i.d. couplings $\{ J_{ij}\}$: this model can be mapped to a
2d quantum single-particle localization problem with random hopping~\cite{Blackman1982, BlackmanPoulter1991, BlackmanPoulterNewInsight, ChoFisher, MerzChalker, ReadLudwigMetal, GruzbergReadLudwigSUSY, MildenbergerDamleGriffiths, Mildenberger, MiddletonHusePRL, ThomasKatzgraber}. 
 Here we focus on the representation of the Ising partition function in terms of an action for Majorana fermions, which contains information about the \textit{entire spectrum} of a pure imaginary Hermitian matrix $\HH$ describing random hopping on the decorated \textit{two}-dimensional ``Kasteleyn'' lattice shown in Fig.~\ref{fig:definitions}~\cite{HurstGreen, kasteleyn, McCoyWuBook}. Here, couplings depend on the inverse temperature, $\beta=T^{-1}$, of the classical Ising model.
This $\HH$ has spectrum $ \{\pm \varepsilon_a(\{J_{ij}\}, \beta) \}$, and the Ising partition function is~\cite{HurstGreen, kasteleyn} 
\begin{equation}\label{eq:ising_gs}
 Z(\{J_{ij}\}, \beta) = 2^{L^2}\left(\prod_{\langle ij \rangle } \cosh(\beta J_{ij}) \right) \prod_{a} \varepsilon_a(\{J_{ij}\}, \beta) .
\end{equation}

For studying transitions induced by the randomness, we use a simple family of distributions of the random couplings throughout this work: unit-variance gaussians parameterized by their mean:\footnote{The commonly studied ``$\pm J$'' distribution, $ P(J) = (1-p) \delta(J - 1) + p \delta(J+1)$, has a similar phase diagram, with $p$ playing the role of $1/\mu$, but extensive ground-state entropy which is important at low $T$ near the SG~\cite{MiddletonHusePRL}. Because $\theta_c > 0$, long-distance physics at criticality in this model is the same as in ours.} $P(J) \propto e^{-(J-\mu)^2/2}$. The phase diagram of the 2d random bond Ising model with this distribution of i.i.d. couplings is shown in Fig.~\ref{fig:phase_diagram}. 
At low $T$ and large $\mu$ the system is a ferromagnet (FM), while at small $\mu$ and $T=0$ it becomes a spin glass (SG). 
Both FM and SG have Edwards-Anderson order, $\lim_{|\bm{x} - \bm{y}| \to \infty} \expect[\langle \sigma_{\bm x} \sigma_{\bm{y}} \rangle^2 ] > 0$. (Here $\langle \cdots \rangle$ denotes a thermal average and $\expect[\cdots]$ a disorder average.)
The FM has non-zero magnetization, $\lim_{|\bm{x} - \bm{y}| \to \infty} \expect[\langle \sigma_{\bm x} \sigma_{\bm y} \rangle] > 0$, while in a SG---here just at $T=0$---this correlator decays exponentially. In the paramagnet 
all correlators decay exponentially.

Each phase is also characterized by a ``stiffness'', defined via the free energy difference $F_{dw}$ between periodic and antiperiodic boundary conditions. The exponent $\theta$ is defined such that the distribution of $F_{dw}/L^\theta$ reaches a nontrivial limit as $L\to\infty$; $\theta$ is the negative RG eigenvalue of $T$ at the controlling $T=0$ fixed point. The 2d ferromagnet has $\theta_{\rm FM} = 1$, i.e. domain wall free energy proportional to length. The 2d SG has $\theta_{\rm SG}<0$~\cite{BrayMoore, McMillanDomainWallRG}: therefore the SG yields to the paramagnet at any $T>0$. 

Increasing either $T$ or $1/\mu$ results in a continuous transition out of the FM. Weak randomness leaves the clean ferromagnetic transition's universality class unchanged, and the subtle physics of this marginally irrelevant RG flow has been studied using free fermionic field theory in terms of moments of the randomness~\cite{DotsenkoDotsenko, ShankarRBIM, Shalaev1994, LudwigHierarchy, NarayanFisher}.
At low $T$ the universality class of the transition changes, and the critical behavior of the FM-SG below the multicritical point is controlled by a distinct $T=0$ fixed point ~\cite{McMillanDomainWallRG, LeDoussalHarrisNishimori}. On this low-temperature critical line, indicated by the bold black line in Figure~\ref{fig:phase_diagram}, $F_{dw} \sim L^{\theta_c}$, with $\theta_c \approx 0.15$~\cite{McMillanDomainWallRG, MelchertHartmann}. 

The flow to $T=0$ on the critical line implies similarities with a thermally stable SG phase (and a finite-$T$ FM-SG transition): all along the low-$T$ critical line we expect two pure Gibbs states.\footnote{Scenarios involving more than two states in 2d are almost precluded by rigorous results~\cite{NewmanStein2dPRL, NewmanStein2dCommMathPhys, Arguin, NewmanStein2025}, and large-scale numerics~\cite{MiddletonPRL1999Thermodynamic, PalassiniYoung, HartmannYoungMetastate}.}
While ferromagnetic correlations decay as a power law---$\expect[\langle \sigma_{\bm x} \sigma_{\bm y} \rangle] \sim |\bm x - \bm y|^{-\eta}$---there is Edwards-Anderson order such that $\lim_{|\bm{x} - \bm{y}| \to \infty} \expect[\langle \sigma_{\bm x} \sigma_{\bm y} \rangle^2 ] > 0$.
This interesting feature---that the 2d low-$T$ critical behavior resembles a higher-dimensional SG phase---has not, to our knowledge, been commented on previously. 
The flow to $T=0$ allows one to numerically study the transition by evaluating ground states, which can be done very efficiently~\cite{Bieche, Barahona, ThomasMiddletonMatching, WeigelDomainWalls} (empirically, in time $\sim L^{ 2.2}$ for $L \lesssim 10^3$~\cite{WeigelDomainWalls}), as discussed further in Appendix \ref{app:kasteleyn_matching}. 

The nature of the low-$T$ transition in the fermionic representation has long been an open issue. We show that its essential feature, which enables understanding in both spaces, is that the flow to $T=0$ in the Ising representation corresponds to a flow to infinite randomness in the fermion representation. This flow to infinite randomness is manifest in the gap in the spectrum of $\HH$ at the critical point, which has a stretched exponential dependence on system size: $\log (\min_a |\varepsilon_a|)^{-1} \sim L^{\psi}$, with $\psi = \theta_c$. 
This correspondence is established using a strong-disorder RG which is exact as $T\to 0$ and asymptotically exact for flows to $T=0$. 
In the quantum representation this RG iteratively diagonalizes $\HH$ using a hierarchy of effective matrix elements whose logarithms become parametrically broadly distributed for small $T$ (either bare or renormalized). This RG is equivalent, on the classical side, to constructing a sequence of Ising Hamiltonians and their ground states, progressively adding in frustration until the desired Hamiltonian $\mathcal{H}$ is reached. 
This type of concrete, asymptotically controlled, realization-specific RG 
has not been achieved (beyond 1d~\cite{IgloiMonthusReview}) for random classical systems;\footnote{See Refs.~\cite{Monthus1, Monthus2, Monthus3} for an infinite-randomness perspective on activated scaling in thermal dynamics associated with $T=0$ fixed points, related to infinite-randomness scaling of spectra of Rokhsar-Kivelson Hamiltonians~\cite{HenleyDynamicsRK}.} previous RGs have dealt with approximations/truncations of {\it distributions} of couplings~\cite{McMillanScalingTheory, BrayMoore,FisherHuseEquilibrium, FisherFunctionalRGInterface, WieseLeDoussalFRGReview, WieseReview, TarjusReview, BricmontKupiainen}.

\subhead{Frustration matchings, and an RG} 
The bridge between classical and quantum pictures is best understood in yet another space: matchings of frustrated plaquettes, which we now describe.
The partition function of the square lattice Ising model is invariant under flipping the signs of four bonds that meet at a spin: these small \textit{gauge transformations} change the locations of negative bonds~\cite{toulouse, FradkinHubermanShenker} but not of \textit{frustrated plaquettes}---squares of four bonds with a negative product $J_{ij}J_{jk}J_{kl}J_{li}<0$---which provide (together with magnitudes $\{|J_{ij}|\}$, and boundary conditions) the gauge-invariant description of a disorder realization. 
Given $\{J_{ij}\}$, any spin configuration $\{ \sigma_i\}$ corresponds to a set of unsatisfied bonds: $\langle ij\rangle $ with $J_{ij} \sigma_i \sigma_j < 0$.  Plaquettes that are frustrated (or not) must comprise an odd (respectively even) number of unsatisfied bonds.
Unsatisfied bonds therefore form strings on the dual lattice that terminate on frustrated plaquettes (together with loops that can be removed by energy-lowering flips of all the enclosed spins). 

Finding Ising ground states is tantamount to minimizing the total weight of unsatisfied bonds (up to boundary conditions discussed in Appendix~\ref{app:kasteleyn_matching}), thus minimizing the total string weight pairing up (\textit{``matching''}) frustrated plaquettes on the dual lattice~\cite{toulouse, Bieche}, where the weight of a string is the sum of $|J_{ij}|$ it crosses. For each pair of plaquettes $p,q \in \mathcal{P}$ (where $\mathcal P$ is the set of all frustrated plaquettes), we define $\zeta_{pq}$ as the minimum---over all paths connecting $p$ to $q$---of the sum of the $|J_{ij}|$ crossed by the path on the dual lattice. 
The ground state corresponds to the perfect (i.e. complete) matching $\MM$ of the plaquettes in $\mathcal{P}$ that minimizes the sum $\sum_{\{p,q\} \in \MM} \zeta_{pq}$.
An example of a ground state, showing the corresponding unsatisfied strings as dashed lines, is shown in Fig.~\ref{fig:definitions}a.

\begin{figure}
 \centering
 \includegraphics[width=\columnwidth]{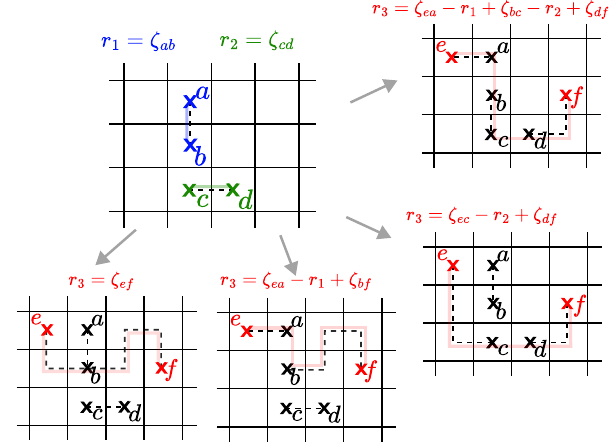}
 \caption{Building up the ground state by adding in frustration. Frustrated plaquettes are labeled by letters. Dashed lines cross strings of unsatisfied bonds that connect frustrated plaquettes. $\zeta_{pq}$ is the cost of directly matching plaquettes $p$ and $q$ in the ground state, calculated as the sum of the magnitudes of the $J$'s along the least costly path between the plaquettes.
In this example, the first two RG steps, $r_1$ and $r_2$, are formed by joining up plaquettes $\{ a,b\}$ and $\{ c,d\}$ along the blue and green paths respectively, indicated in the upper left. When frustrated plaquettes $e$ and $f$ are added, various options are shown for the path (in red) along which the added matching flips the candidate ground state, with new configurations of unsatisfied bonds crossed by the black dashed lines. Of the candidates shown, the red path with the {\it smallest} value of $r_3$ is the one that will be selected in the ground state at this stage of the RG.} 
 \label{fig:rgs}
\end{figure}

Our RG, closely related to an RG of Motrunich, Damle and Huse~\cite{MotrunichDamleHuse}, 
constructs the ground state of $\mathcal{H}$ via a sequence of Hamiltonians $\{\mathcal{H}_{n}\}$ for $n \in \{0, 1, \dots, N_f/2$\}, where $\mathcal H_{n}$ contains $2n$ frustrated plaquettes, with $\mathcal{H}_{N_f/2} \equiv \mathcal{H}$ containing the full set of $N_f$ frustrated plaquettes. 
Each $\mathcal{H}_n$ has the same magnitudes of couplings $\{ |J_{ij}| \}$ but the frustration is built up gradually with $n$, starting from $\mathcal{H}_0$ which is ferromagnetic with couplings $|J_{ij}|$. 

Call $\mathcal{P}_n \subseteq \mathcal{P}$ 
the frustrated plaquettes in $\mathcal{H}_n$, and $2 E( \tilde{\mathcal P})$ the Ising ground-state energy with frustrated plaquettes $\tilde{ \mathcal P}$ 
minus the unfrustrated ground-state energy, $-\sum_{\langle ij \rangle} |J_{ij}|$.
 At each step of the RG, we add two new frustrated plaquettes $p_n, q_n \in \mathcal P$ to make $\mathcal{P}_{n } = \mathcal{P}_{n-1}\cup \{ p_n, q_n \}$. The new plaquettes are chosen to minimize the increment of ground state energy from the old model to the new one, i.e. $\{ p_n, q_n\} = \argmin_{\{p,q\} \subseteq \mathcal{P} \setminus \mathcal{P}_{n-1} } \left[ E(\mathcal{P}_{n-1} \cup \{ p, q \}) - E(\mathcal{P}_{n-1}) \right]$. Define these optimal increments to be $r_n = E(\mathcal{P}_n) - E(\mathcal{P}_{n-1})$, such that the ground-state energy $E_{0,N}$ of $\mathcal{H}_N$ for any $N$ is given by $E_{0,N} + \sum_{\langle ij\rangle} |J_{ij}|= 2 \sum_{n = 1}^{N} r_n = 2 \sum_{\{p,q\} \in \MM_N} \zeta_{pq} $, where the latter sum goes over $p,q$ matched in the ground state of $\mathcal{H}_N$. 
We show in Appendix \ref{app:more_matching} that the increments increase with RG time: $r_n < r_{n + 1}$.

As step $n$ is carried out, the signs of $J_{ij} \sigma_i \sigma_j$ in the ground state (which are negative across unsatisfied strings) flip along \textit{some} path connecting $p_n$ and $q_n$.
Fig.~\ref{fig:rgs} illustrates an example of one RG step. At first the RG typically adds in pairs of frustrated plaquettes that are close to one another (since these have the smallest-weight paths connecting them). 
As the RG progresses, the plaquettes that remain to be matched become more widely spaced and new plaquette pairs are added by flipping small-scale matchings along a longer path.
A natural conjecture is that RG steps at scale $L$ in regions of an infinite system separated by distances $> \mathcal{O}(L)$ are approximately independent. Thus their scaling properties should be similar to those of the (near) final RG steps in a system of size $L$, whose properties we shall focus on presently.


Our RG procedure is very different from finding the ground state of a fixed $\mathcal{H}$ by successively flipping combinations of spins to lower the energy. Instead, we change the {\it Hamiltonian} as the RG progresses, introducing frustration at progressively larger scales, at each step finding the exact ground state via a path that \textit{increases}, by as little as possible, the energy cost due to frustration. While we eventually construct the ground state of the full Hamiltonian $\mathcal{H}$, at earlier steps of the RG the matchings do {\it not} correspond to spin configurations of $\mathcal{H}$: the RG operates in a different space that coincides with the physical spin space only when the procedure is complete. 

 Numerically carrying out all steps of this \textit{gedanken} RG is infeasible---naively requiring time even larger than $\mathcal{O}(L^8)$. 
 However the energy increment from the {\it final} step of the RG can be computed efficiently (see Appendix \ref{app:more_matching}). This $\rmax \equiv r_{N_f/2}$ provides important information about both classical domain walls and the quantum spectrum. 
It can readily be shown (Appendix \ref{app:more_matching}) that $\rmax$ is the largest \textit{decrease} in matching energy that can result from \textit{un}frustrating two plaquettes in $\mathcal{H}$: $\rmax = \max_{p,q \in \mathcal{P}} [E(\mathcal{P}) - E(\mathcal{P} \setminus \{p,q\})]$, and that the optimal removal is $\{p_{N_f/2}, q_{N_f/2}\}$, the final frustrated plaquette pair added in the RG. We will refer to this pair of plaquettes as the \textit{optimized defect} of the matching. 

\subhead{Spectrum of quantum $\HH$} 
 We now turn to the 2d pure imaginary quantum Hamiltonian $\HH$, Fig.~\ref{fig:definitions}b, whose spectrum gives the partition function of the Ising model as Eq.~\eqref{eq:ising_gs}~\cite{HurstGreen, kasteleyn, McCoyWuBook, Blackman1982, BlackmanPoulter1991}. 
We show that the {\it same} sequence $\{r_n\}$ results from an iterative diagonalization procedure for the quantum problem in the $T \to 0$ limit; thus the last step yields the smallest gap in the spectrum.

At $T=0$, all hoppings in $\HH$ are $\pm i$ (since $\lim_{\beta\to\infty}\tanh \beta J_{ij} =\sgn J_{ij}$). It can be shown that $\HH$ then has $N_f$ eigenvectors with zero energy~\cite{Blackman1982, BlackmanPoulter1991}, 
and a basis can be chosen such that one zero-energy mode is localized within each frustrated Ising plaquette. All $4 L^2-N_f$ other eigenmodes have ${\cal O}(1)$ energies as $T \to 0$. As $T$ is increased from zero, the zero modes split into an ``impurity band'' with energy levels $\{ \pm \qe_n\}$. (The index $n$ runs from $1$ to $N_f/2$, in contrast with the $\varepsilon_a$ of Eq.~\eqref{eq:ising_gs} in which $a \in \{1 ,\dots, 2 L^2\}$ labels the entire spectrum of $\HH$.) In order to give a sensible $T\to 0$ limit for the free energy, we must have $\pm \qe_{n} \sim \pm e^{-2\beta R_n}$; this defines the sequence $\{ R_n \}$, ordered so that $R_n < R_{n+1}$.
That the dependence of the low energy quantum states, and hence via Eq.~\eqref{eq:ising_gs} the $T\to 0$ Ising thermodynamics, is solely on the $\{R_n\}$ enables ignoring all prefactors that multiply exponentials. The ground-state energy of the Ising model is therefore $E_0 = - \sum_{\langle ij\rangle } |J_{ij}| + 2 \sum_{n=1}^{N_f/2} R_n$.

For small $T$, $J$-dependent elements of $\HH$ acquire perturbations relative to their $T=0$ values of order $|\sgn J_{ij} - \tanh(\beta J_{ij})| \sim e^{-2\beta|J_{ij}|}$. The effective Hamiltonian, which connects pairs of frustrated plaquettes via virtual hops through unfrustrated plaquettes, involves many such matrix elements of $\HH(T)-\HH(0)$. 
The dominant term {\it directly} connecting $p$ and $q$ can be written in terms of $\zeta_{pq}$ which, as defined earlier, is the sum of the $|J_{ij}|$ crossed by the minimal-weight path between $p$ and $q$.
The impurity-band spectrum of $\HH$ in the large-$\beta$ limit is then given by the spectrum of the $N_f\times N_f$ antisymmetric matrix $\boldups $ with elements $\Upsilon_{pq} =-\Upsilon_{qp}= ie^{-2\beta \zeta_{pq}}$. (The assignment of signs depends in a complicated way on the locations of the plaquettes, but this is of no consequence at large $\beta$.)

Consider a sequence of effective Hamiltonians $\{\boldups_n\}$ corresponding to the Ising Hamiltonians $\{\mathcal{H}_n \}$, such that $\boldups_n$ is the $2n \times 2n$ matrix containing elements $\sim \pm i e^{-2\beta \zeta_{pq}}$ for all pairs $p, q\in \mathcal{P}_n$.
The connection between the classical RG and the quantum procedure is that \textit{the sequences $\{r_n\}$ and $\{R_n \}$ are identical,} with the first sequence coming from the gradual addition of frustration to the Ising Hamiltonian, and the second sequence coming from the log quantum energies. For $\beta \to \infty$, plus/minus energy pairs of impurity-band wavefunctions are localized on plaquettes $p_n, q_n$. We show the equivalence between classical and quantum sequences $\{r_n\}$ and $\{R_n\}$ inductively in Appendix \ref{app:more_fermion}; a rough argument follows.

In going from $\boldups_{n-1}$ to $\boldups_n$ we append two rows and columns containing matrix elements connecting plaquettes $p_n$ and $q_n$ (the same plaquettes as were added at this step in the matching RG) to the plaquettes in $\mathcal{P}_{n-1}$. 
The effective $2\times2$ Hamiltonian on the $p_n,q_n$ space receives contributions from all virtual excursions through $\boldups_{n-1}$, alternating between $e^{-2\beta \zeta}$ elements as numerators and $e^{-2\beta r}$ elements as denominators in perturbation theory. Each such term corresponds to a possible change in matching (red lines in Fig.~\ref{fig:rgs}). The alternating sums of $\zeta$'s and earlier $r$'s thus correspond to possible increments of the Ising ground state energy. The largest such alternating product, which dominates as $\beta \to \infty$, is $e^{-2\beta r_n}$: this corresponds to the smallest energy increment in the Ising representation.\footnote{With $\{ J_{ij}\}$ all equal to $\pm 1$, 
a similar degenerate perturbation theory, with $\{ r_n\}$ integers corresponding to orders in perturbation theory, was discussed by Blackman and Poulter~\cite{Blackman1982, BlackmanPoulter1991, BlackmanPoulterNewInsight}.}

\begin{figure}
 \centering
 \includegraphics[width=\columnwidth]{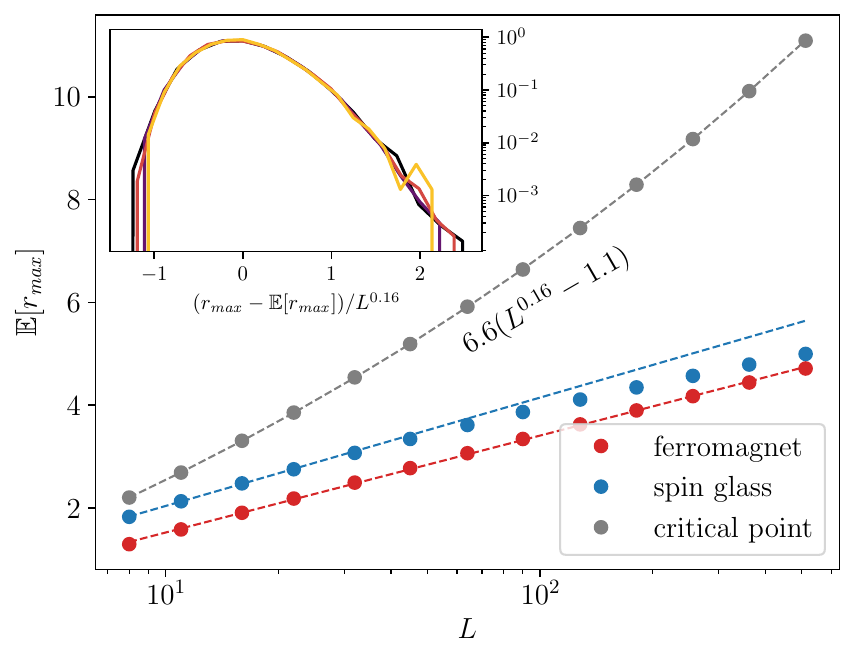}
 \caption{Scaling of the optimized defect energy $\rmax$ with $L$. In the spin glass ($\mu=0$), $\rmax$ grows more slowly than $\log L$. In the ferromagnet ($\mu=1.5$), $\rmax$ grows as $\log L$. (Blue and red dashed lines are $\sim\log L$.) At the critical point ($\mu=1.0307$), $\rmax$ grows as $L^{0.16}$ 
 with a constant correction to scaling (dashed black line fit to the gray sequence of points). Error bars are smaller than marker sizes.
 The inset shows, at criticality, the scaled distribution of $\rmax$ minus its mean, for several $L$ between $64$ and $512$.
 Averages are over at least $3\times 10^4$ disorder realizations. }
 \label{fig:rmax_L}
\end{figure}

We now discuss the scaling behavior on the critical line and in the two phases, focusing first on the information in the last step of the RG.
The smallest splitting in the quantum spectrum, $\emin \equiv \qe_{N_f/2}$, is simply related to the optimized defect energy $\rmax \equiv r_{N_f/2}$ as $\rmax = \lim_{\beta \to \infty} (\frac{1}{2\beta}\log \emin^{-1})$. We used the fast algorithm in Appendix \ref{app:more_matching} to study how the distribution of $\rmax$, and therefore of $\log \emin$ in the quantum problem, scales with $L$: results are shown in Fig.~\ref{fig:rmax_L}.

(It should be emphasized that we deduce spectral properties of $\HH$ without ever diagonalizing it. Diagonalization would be costly---especially due to precision demanded at low $T$ by exponentially small eigenvalues---and, importantly, conceptually inferior for the critical point: we instead work directly in the $T\to 0$ limit, where the physics lives in the log quantum energies.)

\subhead{Critical point} At the critical point $\mu = \mu_c$, both the mean and width of the distribution of $\rmax$ grow faster than $\log L $. Since $\emin \sim e^{-2\beta \rmax}$ for large $\beta$, there is an infinite-randomness singularity in the spectrum of the critical $\HH$ in the $T\to 0$ limit.\footnote{Broadening of the distribution of $\{R_n\}$ at criticality was hinted at in Ref.~\cite{BlackmanPoulterNewInsight}.} 
We now argue for the scaling $\rmax \sim L^\psi$, with $\psi = \theta_c$, the stiffness exponent at the FM-to-SG critical point.

In an infinite-lattice ground state, consider minimal-energy flips restricted to, say, contain a given size-$L$ box while staying within a size-$2L$ box outside it. 
This is one definition of \textit{droplets} at scale $L$~\cite{FisherHuseEquilibrium}.
One often assumes approximate statistical independence of scale-$L$ and $2L$ droplets, and that the scale-$L$ droplets associated with two regions are independent only when they are at least order-$L$ apart. 
At the critical point, we expect the distribution of droplet energies to scale with $L^{\theta_c}$.
Crucially, if we divide a size-$L$ droplet into an order-one number of partial boundaries, each will contribute an energy that scales with $L^{\theta_c}$, but with a prefactor that has an order-one probability of being negative. Optimized defects at scale $L$ then go from one frustrated plaquette to another along paths that are like the surfaces of droplets, and indeed along unfrustrating paths when possible, maximizing unfrustration at each scale. 
Thus we get a coherent sum over scales: $\expect[\rmax] \sim \sum_{n=0}^{\log_2 L} c_n (2^n)^{\theta_c}$. (Such summed-over-scales quantities are commonplace in critical phenomena, e.g. free energies at usual critical points.) The sum is dominated by the order-$L^{\theta_c}$ term, but the smaller scales contribute an order-one (presumably leading) correction to scaling.

A fit to $\expect[\rmax] = A(L^{\theta_c} + BL^\phi)$ with $\theta_c=0.15$ gives $B<0$ and $\phi \approx 0.01 $ (with error at least as big as mean), consistent with a constant leading correction to scaling.
 An alternative fit of the form $\expect[\rmax] = a( L^{\psi} - b)$ (also with three parameters) yields $a = 6.6 \pm 0.1$, $b = 1.065 \pm 0.005$, and $\psi = 0.16 \pm 0.01$, in agreement with $\theta_c$ within error bars. 
The $n$ dependence of $\sgn(c_n)$ in the above sum over scales (assuming all magnitudes $\mathcal{O}(1)$) is nontrivial: we expect $c_{\log_2 L}>0$ since the optimization must be dominated by scale $L$, but possibly $c_0 < 0$ as at scale $1$ the optimized defect may be more likely to frustrate than to unfrustrate, due to the (``bare'') coupling distribution having $\expect[J]>0$; the sign of $c_0$ is certainly non-universal and could be tuned by the distribution of couplings. 
Note that one also expects $o(1)$ corrections such as from $(L+\text{cst.})^{\theta_c} \approx L^{\theta_c} +{\cal O}( L^{\theta_c-1})$, much smaller for small $\theta_c$, and possible non-trivial corrections to scaling. 

The infinite randomness nature of the critical point is reflected in the distribution of $\rmax$. 
The inset of Fig.~\ref{fig:rmax_L} shows the distribution of $(\rmax-\expect[\rmax])/L^{\psi=0.16}$ exhibiting a scaling collapse over a range of $L$ (this eliminates the constant correction of the mean; the \textit{variance} of the constant part is expected to be small, roughly because optimization over many scales pushes into tails of distributions). 

\subhead{Asymptotic validity of RG at criticality} 
Our direct analysis is of the large-$\beta$ limit of the spectrum (and the classical ground states). In this limit, the distribution of $\emin$ is parametrically broad in $\beta$ on a log scale at all $L$ since $\emin \sim e^{-2\beta\rmax}$. Correspondingly, the RG constructing the ground state is always controlled, provided we take the $T \to 0$ limit before $ L\to \infty$. At criticality, however, the distribution of $\rmax$ broadens without bound \textit{as a function of $L$}. This implies \textit{asymptotic} correctness of the RG: infinite randomness behavior obtains at large scales along the whole low-$T$ critical line. 
A matrix $\Upsilon$ in the basis of would-be zero modes can in principle be constructed exactly by integrating out all other modes (as we have free fermions). When $\beta < \infty$, the elements of $\Upsilon$ do not have a parametrically broad distribution. However the iterative diagonalization RG of $\Upsilon$ can still be run, though it is no longer associated with an Ising Gibbs state. If, at some $\mu \in (0, \infty)$, the distribution of logarithms of elements at scale $L$ in $\Upsilon$ broadens with $L$---which we expect it to, by the stability of the $T=0$ fixed point---the RG becomes asymptotically controlled even though it makes errors at short scales. Thus large RG scale at criticality plays the role of large $\beta$. 

\subhead{Ferromagnet} 
Deep in the FM phase, negative bonds are rare and thus typically produce tightly bound pairs/clusters of frustrated plaquettes. The RG matches these plaquettes initially, so the only ones unmatched towards the end of the RG are far-separated pairs produced by long strings of negative bonds which are exponentially rare in their length. Hence $\expect[\rmax] \sim \log L$.
Correspondingly, the quantum energies decrease as a power of length scale, $\emin \sim L^{-z}$, characteristic of a ``Griffiths phase'' with continuously varying exponent $z$. This phase was discussed in Refs.~\cite{MotrunichDamleHuse, MildenbergerDamleGriffiths}. (See in Appendix \ref{app:1d} comments on a one-dimensional problem that substantially shares this physics.) 

The flow to $T=0$ of the FM can be understood in terms of RG trajectories: at scale $L \gg 1$ the distribution of $\zeta$ splits into two classes: (1)
typical distances between the rare remaining frustrated plaquettes and their remaining nearest neighbor scale as $\log L$ and govern $\rmax$, while (2) typical distances separating such pairs scale as $L$
and make the domain-wall free energy proportional to $L$. Elements in $\Upsilon$ of the former class are asymptotically larger than the latter, and so the RG is still controlled in the FM Griffiths phase~\cite{FisherTFIM}.

\subhead{Spin glass/paramagnet}
 Deep in the SG, there is little clustering of frustrated plaquette pairs, so rare defects with linear-in-length energy gain are even rarer than exponential. Indeed we observe $\rmax = o(\log L)$. The weak growth with $L$ is perhaps from small-scale optimization of the ends of the defect over polynomially many locations; contributions from large scales vanish as $L\to \infty$ since $\theta_{\rm SG} < 0$. 

The low-$T$ flows of the near-critical paramagnet ($\mu = \mu_c - \delta$) are interesting: at the scale of the correlation length associated with the critical point, $\xi\sim \delta^{-\nu}$, the bare temperature $T$ has renormalized down to $ T \delta^{\nu \theta_c}$. Then, governed by the flow away from the SG fixed point, it starts to grow, and becomes order-one at length scale $ (T\delta^{\nu\theta_c})^{-1/|\theta_{\rm SG}|}$. Our RG is expected to be controlled out to this scale, at which it predicts its own demise, as distributions are no longer parametrically broad.

\subhead{Speculation} It is tempting to conjecture that there are general connections between random systems such as finite-$T$ spin glasses controlled by classical zero-temperature RG fixed points on the one hand and infinite-randomness fixed points---in an appropriate ``space''---on the other. 
Can one use optimized defects at various scales, explicitly constructed in our RG, as the basis for phenomenological theory, complementary to droplets~\cite{FisherHuseEquilibrium, FisherHuseNonequilibrium}?
Does sensitivity of RG trajectories to small changes in couplings/temperature have relations to disorder/temperature chaos in spin glasses?
Can such concrete asymptotically controlled RGs be formulated for other $T=0$ fixed points, in particular the 3d SG? 
If yes, can many-state scenarios provide precise conjectures about the character of these infinite-randomness fixed points?

\subhead{Acknowledgments}  We thank the Stanford Sherlock computing cluster for resources.
We are grateful to Kedar Damle for discussions and correspondence, and to Nicholas O'Dea for careful comments on the manuscript.
This work was supported in part by the National Science Foundation under Grant DMR-2000987 (AP) and PHY-2210386 (DSF), and by grants from the NSF (DMS-2235451) and Simons Foundation (MPS-NITMB-00005320) to the NSF-Simons National Institute for Theory and Mathematics in Biology (AM).

\bibliography{references}

\onecolumngrid
\newpage 
\appendix

\begin{spacing}{1.2}

\section{Kasteleyn matching}\label{app:kasteleyn_matching}

\begin{figure*}
 \centering
 \includegraphics[width=0.4\linewidth]{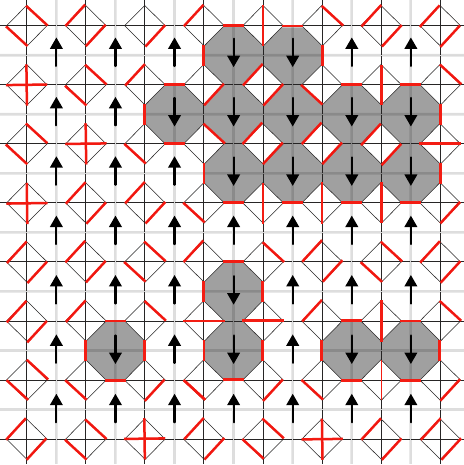} \qquad
 \includegraphics[width=0.4\linewidth]{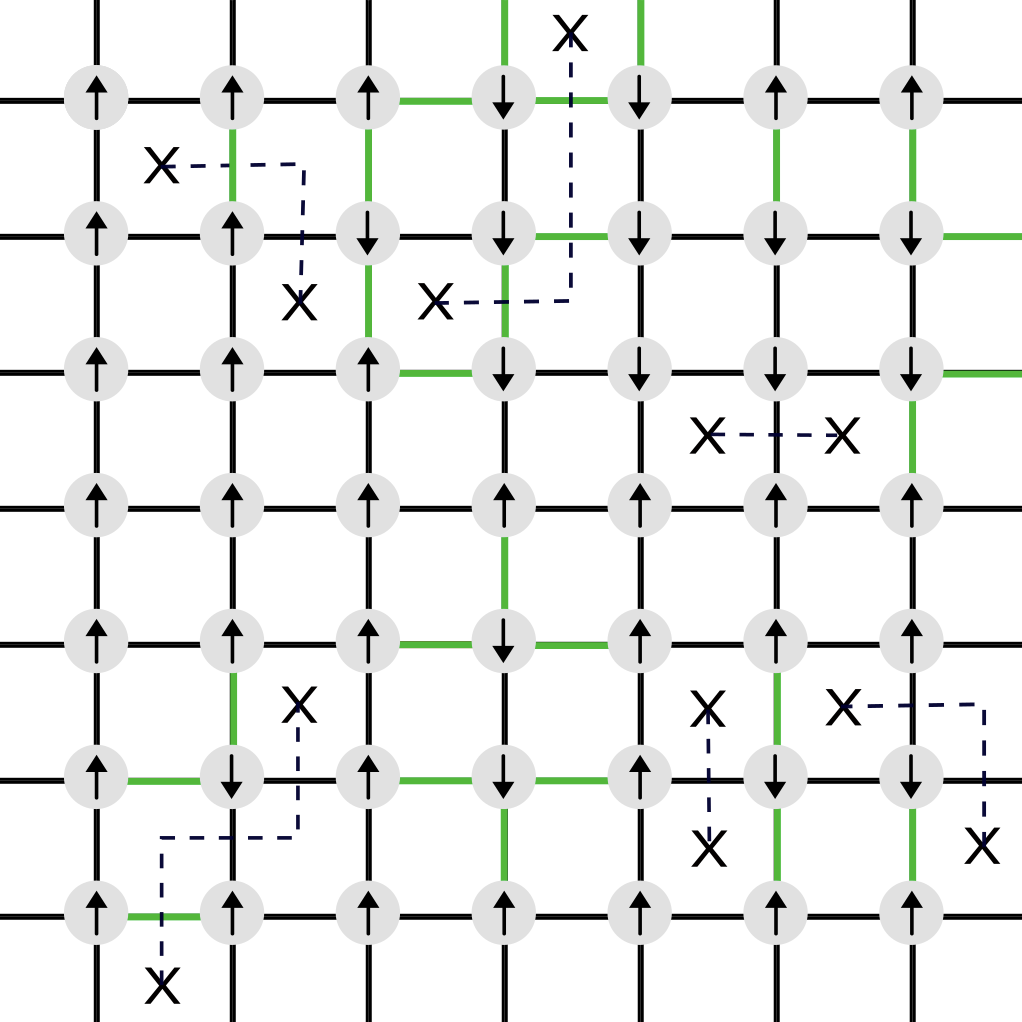}
 \caption{ Left: the dual Kasteleyn lattice and, in red, a perfect matching on it, which maps to the displayed spin configuration. Each four-site clique (Kasteleyn city) corresponds to a plaquette. The grey regions, containing spins that are overturned relative to the uniform state, are surrounded by domain walls which are closed polygons on the dual lattice that also define dimer-hosting intercity bonds. \quad Right: the primal Ising lattice (solid lines). Green bonds are antiferromagnetic, and their positions define the gauge invariant locations of the frustrated plaquettes (crosses). The same state is represented by strings of unsatisfied bonds (dashed) that in the ground state constitute a minimal-weight perfect matching of the frustrated plaquettes. }
 \label{fig:kasteleyn} 
 
\end{figure*}

In this Appendix we explain the map between the Ising ground state problem and the efficiently solvable combinatorial optimization problem of minimum weight perfect matching (MWPM) which we used for our numerics~\cite{ThomasMiddletonMatching}. 
Matchings of frustrated plaquettes (which we will refer to as Toulouse matchings and define more concretely presently) were informally discussed in the main text; the MWPM map we will deal with first is, however, different.

First, for the discussion here and in the following appendices, it is useful to define some terms precisely. We will work with finite undirected graphs $\GG=(\VV,\EE)$ with vertices $\VV$ and edges $\EE$. Edges carry weights given by a function $w:\EE\to \mathbb{R}$. A \textit{perfect matching} on $\GG$ is a subset of edges $\MM \subseteq \EE$ such that for every $v \in \VV$ there is exactly one $e \in \MM$ adjacent to $v$. A \textit{partial matching}, meanwhile, is an edge subset $\tilde{\MM} \subseteq \EE$ such that for every $v \in \VV$ there is at most one $e \in \tilde{\MM}$ adjacent to it. 
Thus a partial/perfect matching is an imperfect/perfect dimer cover of $\GG$ respectively. We will often loosely talk of vertices in a matching $\MM$: this will mean the vertices $v\in \VV$ that are adjacent to some edge in the matching $\MM$. 
For either a perfect or a partial matching, we define the weight of the matching $\mathcal{W}(\MM) = \sum_{e \in \MM} w(e)$. The MWPM weight of $\GG$ is $E(\GG) = \min_\MM \mathcal{W}(\MM)$, with the MWPM being $\MM_0 \in  \argmin_{\MM} \mathcal{W}(\MM)$ (there may be multiple MWPMs). 

MWPM is a well-studied optimization problem, with efficient polynomial-time algorithms for general graphs~\cite{Edmonds_1965, kolmogorov2009blossom}. 

The lattice $\GG$ on which the MWPM problem corresponding to the square-lattice Ising model lives has exactly the same structure as the lattice on which the quantum hopping problem $\HH$ lives (Fig.~\ref{fig:definitions}), however now every four-site city will correspond to an Ising \textit{plaquette}, i.e. a site of the dual lattice. Intra-city weights are all zero, and an inter-city edge crossing Ising bond $ij$ carries weight $J_{ij}$---positive or negative. We will call this graph $\GG_K$, the dual Kasteleyn lattice~\cite{kasteleyn}; see Fig.~\ref{fig:kasteleyn}.

The similarity of lattices is no coincidence: the Ising partition function can be written either as a sum over closed polygons $L$ on the primal square lattice, $\sum_L\prod_{ij\in L} \tanh (\beta J_{ij})$, or on the dual lattice, $\sum_{\tilde L}\prod_{ij\in \tilde L} e^{-2\beta J_{ij}}$. While the expression in Eq.~\eqref{eq:ising_gs} that comes from the spectrum of $\HH$ is equivalent to the former expansion~\cite{HurstGreen, kasteleyn, McCoyWuBook}, the latter is the Kramers-Wannier dual (one could equally well define $\HH$ with nontrivial elements equal to $\pm i e^{-2\beta J_{ij}}$) and is useful for determining the spin configuration that minimizes the energy~\cite{ThomasMiddletonMatching} because the configuration of closed polygons that dominates the expansion at low temperature is the outline of clusters of spins overturned (relative to the all-up state) in the ground state.

Each perfect matching $\MM$ of $\GG_K$ corresponds to a global-flip-related pair of Ising spin configurations $\pm \sigma^\MM_i $, such that every bond in a domain wall of $\sigma^\MM$ (namely, every bond $ij$ such that $\sigma^\MM_i \sigma^\MM_j=-1$) crosses an edge in $\MM$.
This construction is illustrated in Fig.~\ref{fig:kasteleyn}, where the edges in red are matched. Closed polygons on the dual Ising lattice are outlines of overturned clusters in the spin configuration. These closed polygons can be ``completed'' to a perfect matching of the Kasteleyn lattice, which is unique modulo rearrangements on cities on which no external dimers are incident---each such city contributes a factor of $3$ to the degeneracy of the MWPM. 
The total weight of $\MM$ is equal to the Ising energy of the equivalent configuration relative to the all-up configuration, $\mathcal{W}(\MM) = -\sum_{\langle ij \rangle}J_{ij} (\sigma^\MM_i \sigma^\MM_j -1) $.
Thus finding Ising ground states is equivalent to finding a MWPM of the appropriately weighted dual Kasteleyn lattice $\GG_K$.

We must now consider boundary conditions. For planar Ising graphs---which include those with open boundary conditions but also cylinders as they can be ``wrapped" in the plane---the MWPM as stated above determines the ground state, with the Kasteleyn lattice terminated on an open boundary such that there is one row of cities outside the last row of spins. The MWPM on a torus, on the other hand, determines the ``extended ground state'',
 defined as the configuration that has the minimal energy among the $4$ $(=2^2)$ choices of periodic/antiperiodic boundary conditions in either direction~\cite{ThomasMiddletonMatching}. The extended ground state is a valid spin configuration only for the energetically optimal choice of these four possible boundary conditions. 
 That is, in general one has to insert appropriate non-contractible domain walls (by flipping signs of a path of bonds in the appropriate direction) in order to make the MWPM on the torus correspond to a valid spin configuration. Given a MWPM on a torus, one can determine which choice of boundary conditions corresponds to this extended ground state by counting the number of domain walls separating up and down spins in each of the two directions: an odd number of domain walls indicates that antiperiodic boundary conditions are energetically favored along that direction.

For completeness, let us define the problem of MWPM on the graph of frustrated plaquettes~\cite{toulouse, Bieche, Barahona}, which we have used at various points in the main text. The graph in question, which we shall call the \textit{Toulouse graph} $\GG_T$, is the \textit{complete} graph of frustrated plaquettes: $\VV= \mathcal{P} = \{ \text{plaquettes } p = \prescript{i}{l}\square^j_k \mid J_{ij}J_{jk}J_{kl} J_{li} < 0 \}$, and $\EE = \left\{ \{p,q\} \mid p,q \in \VV,\ p \neq q \right\}$. Defining informally paths $P_{pq}$ on the dual lattice between plaquettes $p$ and $q$, weights on the Toulouse graph are given by $\zeta_{pq} = \min_{P_{pq}}\left(\sum_{\langle ij\rangle \text{ crossing } P_{pq}} |J_{ij}|\right)$. MWPM on $\GG_T$ yields Ising ground states, with unsatisfied strings on the dual lattice connecting a matched pair of plaquettes $pq$ along the path $P_{pq}$ that produced weight $\zeta_{pq}$. 
The Ising ground-state energy is then given by $ 2 E(\GG_T) -\sum_{\langle ij \rangle} |J_{ij}| $.

Degeneracies in the MWPM Toulouse graph problem are more severe: even if the Ising ground state pair is unique, we get degeneracies whenever two unsatisfied strings cross each other, since such crossings can be resolved to produce several different matchings on $\GG_T$. Boundary conditions are also somewhat more subtle: while periodic boundaries again give extended ground states as defined above, to deal with open boundaries one has to augment the graph $\GG_T$, allowing plaquettes to match with the boundary, corresponding to unsatisfied bonds on the boundary~\cite{Bieche}. We will not discuss these issues associated with Toulouse matching: our numerics will always use the dual Kasteleyn lattice as it is much faster. However for descriptions of the RG involving Toulouse matching, one may imagine either large systems with open boundaries but all frustrated plaquettes far away from the boundaries, or ``magnetized'' boundaries carrying infinite ferromagnetic $J$'s, so that boundary bonds are never unsatisfied.

\section{Domain wall numerics}\label{app:dw}

Using the map to MWPM on the dual Kasteleyn lattice defined in Appendix~\ref{app:kasteleyn_matching}, we use efficient evaluation of ground states to precisely determine the location of the $T=0$ FM-SG critical point and measure exponents associated with domain-wall excitations at this transition. Our results are in agreement with those of Refs.~\cite{AmorusoHartmann, Picco, MelchertHartmann}, 
which studied some of the same observables with MWPM of the Toulouse type, using an older version of algorithms for MWPM.

\begin{figure}
 \centering
 \includegraphics[width=.45\columnwidth]{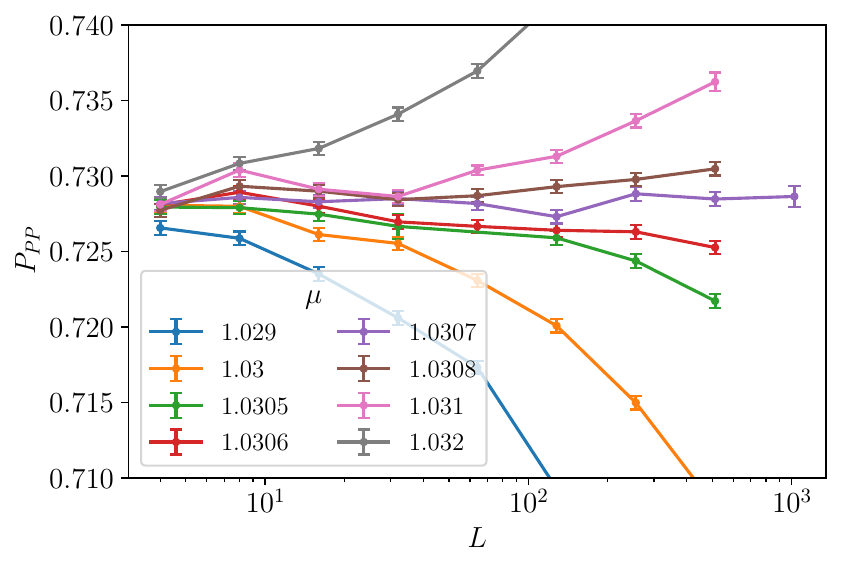}
 \caption{ The probability $P_{PP}$ of the extended ground state selecting fully periodic boundary conditions as a function of system size $L$. Each data point is averaged over between $2\times10^5$ and $10^6$ disorder realizations. }
 \label{fig:critical_point}
\end{figure}

\subsection{Locating the critical point}
There are numerous ways to locate the $T=0$ critical point of the model as a function of $\mu$. 
Here, we use $\PPP$, defined as the probability of selecting periodic boundary conditions in both directions in the extended ground state on the torus. A similar quantity was used in Ref.~\cite{ThomasKatzgraber} to define the finite-temperature critical line. It follows from the definitions in Appendix \ref{app:kasteleyn_matching} that $\PPP$ is the probability that relative domain walls along all non-contractible loops cost positive energy, i.e. $\PPP = \prob[E_{0,\it PP} < E_{0,\it AP}, E_{0,\it PA}, E_{0,\it AA}]$.
In the FM phase, the distribution of the domain wall tension $(E_{0,\it AP}- E_{0,\it PP})/L$ concentrates to a delta function as $L\to\infty$, so that $\PPP \to 1$.
On the other hand, in the SG, at large $L$ the distribution of $ E_{0,\it AP}- E_{0,\it PP} $ becomes symmetric about zero (and scales with $L^{-|\theta_{\rm SG}|}$ as $\theta_{\rm SG}<0$), and therefore $\PPP$ approaches $1/4$. 
At the critical point, the mean domain wall cost diverges with system size as $L^{\theta_c}$, and the distribution of $(E_{0,\it AP}- E_{0,\it PP}) /L^{\theta_c}$ is expected to approach a universal 
 form. This distribution has nonzero support on negative values, but is asymmetric as at all scales the system ``knows" that $\expect[{J}] >0$. Therefore, at criticality,
$\PPP$ goes to a universal number 
between $1/4$ and $1$ as $L\to\infty$ in an $L\times L$ system.
In Fig.~\ref{fig:critical_point}, by inspecting the curves $\PPP(L)$ at various $\mu$, we can identify the critical point as a separatrix at \begin{equation}
 \mu_c = 1.0307 \pm 0.0001, 
\end{equation}
in agreement with Refs.~\cite{MelchertHartmann, Picco}, with
\begin{equation}
 \PPP(\infty) = 0.7285 \pm 0.0010.
\end{equation}

\subsection{Domain walls on cylinders}

\begin{figure*}
 \centering
 \includegraphics[width=.45\columnwidth]{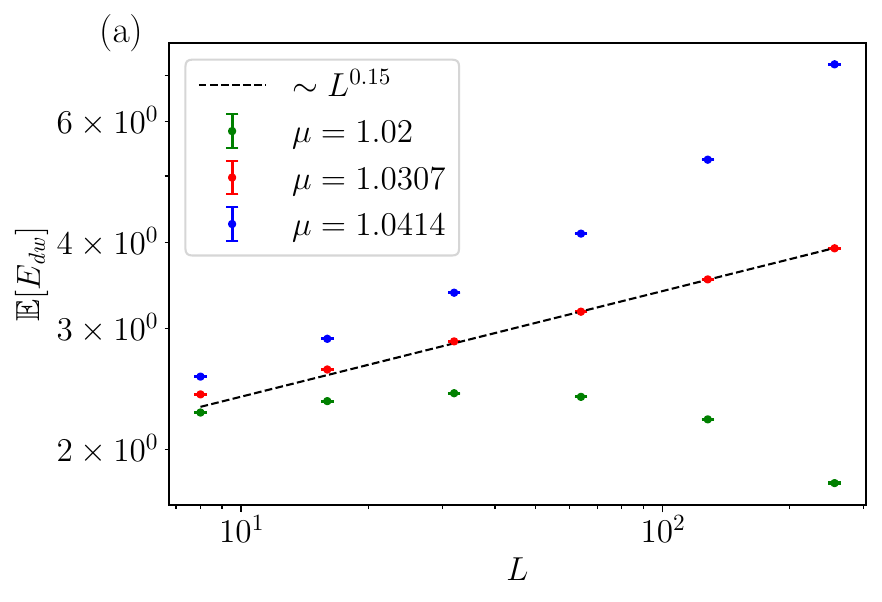}
 \includegraphics[width=.45\columnwidth]{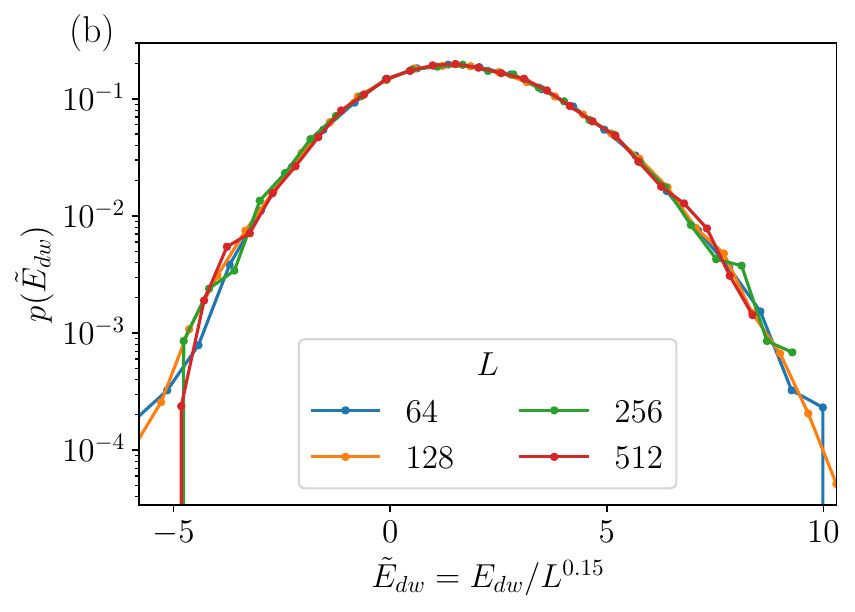}
 \includegraphics[width=.45\columnwidth]{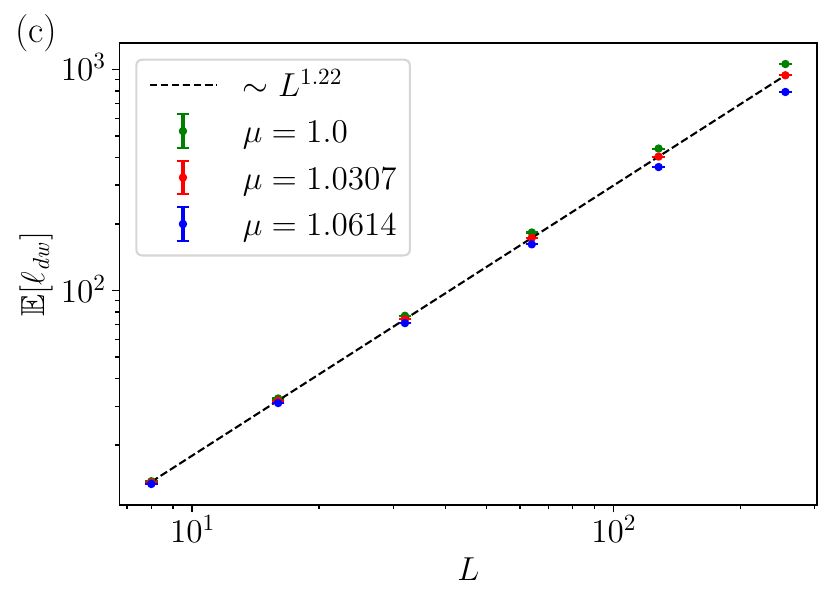}

 \caption{ (a) The mean domain wall energy as a function of $L$ on either side of criticality, as well as at the critical point $\mu_c \approx 1.0307$. The dotted line---a fit to the data at $L\geq 64$---is proportional to $L^{\theta_c}$, with $\theta_c$ given by Eq.~\eqref{eq:theta_c}. Error bars shown are the standard error. (b) the distribution of domain wall energies at the critical point, $\mu = \mu_c$, scaled by $L^{\theta_c}$. This scaling distribution at the critical point is expected to be a universal function independent of disorder distribution. (c) the mean domain wall length as a function of $L$ for $\mu$ on either side of the critical point, as well as at the critical point. The dotted line is proportional to $L^{{\df}_c}$, with ${\df}_c$ given by Eq.~\eqref{eq:dfc}. Numerics for (a) and (c) were carried out with $10^6$ disorder realizations, and for (b) with $3\times 10^4$ disorder realizations. 
 }
 \label{fig:dw_en_length_numerics} 
\end{figure*}

There are various possible measures of ``stiffness'' at the critical point, and hence many ways to diagnose the character of low-energy excitations, all of which are expected to have the same scaling with system size---at least if there are, as now almost proved in 2d~\cite{NewmanStein2025}---only two infinite system ground states. Here we consider the energetics and geometry of domain walls introduced by a relative flip of boundary conditions. While the torus was appropriate for calculating $\PPP$, finding ground states with specified boundary conditions requires planarity, and thus we now use $L\times L$ cylinders, i.e. boundary conditions periodic in one direction and free in the other. 

We calculate the domain wall energy by taking the difference in ground-state energies (which we call $E_{\it dw}$) between the original configuration and one that differs from it by a flip in bond signs along a path on the dual lattice from one free end of the cylinder to the other. These two ground states differ by a pair of relative domain walls: one is the path of flipped bonds, with zero contribution to the energy, and the other is free to optimize its location to minimize energy. The results are shown in Fig.~\ref{fig:dw_en_length_numerics}.

The mean domain wall energy $\expect[E_{\it dw}]$ grows linearly with $L$ in the FM, and decays to zero in the SG. At the critical point $\mu = \mu_c$, $\expect[E_{\it dw}]$ grows as $L^{\theta_c}$. We find 
from our numerics that
\begin{equation}\label{eq:theta_c}
 \theta_c = 0.15 \pm 0.01,
\end{equation}
in agreement with Ref.~\cite{MelchertHartmann}.
As expected, the entire distribution of domain wall energies at the critical point scales with $L^{\theta_c}$, revealing a universal scaling function, asymmetric with positive mean. Around a fifth of disorder realizations at $\mu=\mu_c$ have $E_{\it dw}<0$. 

Note that, in contrast to the optimized defect energies discussed in the main text, energies of domain walls on the cylinder seem not to have an order-one correction to scaling. This is because the domain wall on the cylinder at criticality does not ``know" whether it should be there or not except on the largest scales---reflected in its energy being negative with substantial probability. Locally near its ends, the statistics of the interactions are almost the same as with periodic boundary conditions. Thus small segments of the domain wall near the boundary are not atypical and do not give rise to ${\cal O}(1)$ contributions to its energy. In contrast, the optimized defect is created by taking out two frustrated plaquettes, and the locations of plaquettes are not independent. Thus, near the end points, the statistics of the couplings are atypical and the contribution of the parts near the endpoints of the induced part of a domain wall are systematically non-zero.

We also plot the mean of the length $\ell_{\it dw}$ of domain walls as a function of $L$ at $\mu = \mu_c$. Domain walls at the critical point have wandering exponent ``$\zeta = 1$'', crucially with overhangs on all scales and a nontrivial fractal dimension, $\expect[\ell_{\it dw}] \sim L^{{\df}_c}$. We find, in agreement with Ref.~\cite{MelchertHartmann}, that the fractal dimension is 
\begin{equation}\label{eq:dfc}
 {\df}_c = 1.225 \pm 0.002.
\end{equation}
The fractal dimension of domain walls in the FM is $1$---transverse fluctuations scale like $L^\zeta$ with $\zeta = 2/3 < 1$~\cite{HuseHenleyFisherRespond}---while in the SG there is again a nontrivial fractal dimension $d_{f,{\rm SG}} \approx 1.27$, slightly higher---more than apparent error bars---than ${\df}_c$~\cite{WeigelDomainWalls}.

We have also studied the scaling of the length of optimized defects at criticality with $L$ (not shown), and found that it is consistent with $ L^{{\df}_c}$, which provides further evidence for the proposition that the paths taken by optimized defects can be thought of as partial domain walls or droplet boundaries.

\section{More on matchings}\label{app:more_matching}

It will be useful to introduce some more notation. 

Given a partial matching $\tilde{\MM}$ of $\GG =(\VV,\EE)$, an \textit{augmenting path} (AP) for $\tilde{\MM}$ is a path $(v_1, v_2, \cdots, v_{n-1}, v_n)$ that starts and ends on unmatched vertices ($v_1$ and $v_n$ are not matched in $\tilde{\MM}$, while all other $v_i$ are) and alternates between matched and unmatched edges: $\{v_1 ,v_2\} \notin \tilde{\MM}, \{v_2 ,v_3\} \in \tilde{\MM}, \dots, \{v_{n-1} ,v_n\} \notin \tilde{\MM} $. The alternating sum of edge weights along an AP gives the increase in the cost of the partial matching incurred by flipping the dimer occupancy of each of the edges in the path and thereby adding the two endpoints to the partial matching.
Note that single edges between unmatched vertices are also APs---their alternating sum is just their edge weight. Finding APs is a way of constructing a perfect matching on any graph, and is used in the Blossom algorithm for MWPM~\cite{Edmonds_1965, kolmogorov2009blossom}.

Note, then, that our RG in the Ising problem is tantamount to a sequence of partial matchings on the complete Toulouse graphs $\GG_{T,n}$ whose vertex sets are respectively $\mathcal{P}_n$, such that the increment in the MWPM weight incurred at each step is minimized; these increments are the $r_n$.

Degeneracy in the MWPM of $\GG_T$---which results from zero-weight alternating loops---does not correspond to Ising degeneracy, and while it impacts the choice of partial matchings on $\GG_T$ throughout the RG, it does not result in degeneracies of $r_n$, or in ambiguities regarding the order in which plaquettes are added. Therefore in the following we can have in mind some specific order of partial matchings on $\GG_T$ as the RG trajectory.

\subsection{Monotonicity of $r_n$}

We claimed in the text that for all steps $n$ of our RG, $r_n < r_{n + 1 }$, and we shall now (informally) show this.

Assume for the sake of contradiction that we make an AP with increment $r_n$, which creates access to another AP with alternating sum $r_{n +1}<r_{n}$.
The APs must overlap, otherwise the AP corresponding to $r_{n+1}$ would have been formed before the AP corresponding to $r_{n}$. Let us say that the contribution to $r_{n}$ from the overlapping region is $\rho$, which is somewhere in the middle of both APs. Then $r_{n} = d_1+\rho+d_2$ where $d_1$ and $d_2$ are the other two segments of the first AP. Similarly $r_{n+1} = \tilde d_1 - \rho +\tilde d_2$ where $d_1,d_2,\tilde d_1,\tilde d_2$ are non-overlapping. 
Note the difference in signs of $\rho$, since the alternating sum along the shared segment will be negated when we form the second AP. In order for $r_{n}$ to be made first, then, 
\beq d_1+\rho+d_2<\tilde d_1+d_1,\quad\text{and}\quad d_1+\rho+d_2<\tilde d_2+d_2,\eeq
since both paths corresponding to $\tilde d_1+d_1$ and $\tilde d_2+d_2$ are also valid augmentations. Therefore $\tilde d_1>\rho+d_2$ and $\tilde d_2>\rho+d_1$. In order to have $r_{n + 1}<r_{n}$, or equivalently $\tilde d_1 -\rho +\tilde d_2<d_1+\rho+d_2$, we must have
\beq \rho + d_2 + \rho+ d_1 - \rho <d_1+d_2+\rho\implies 0<0,\eeq
which is a contradiction. 
For a continuous distribution $P(J)$, which we always use, $r_n < r_{n+1}$ with probability one.

We note at this point that the sequence of APs that the Blossom algorithm \cite{Edmonds_1965, kolmogorov2009blossom} for MWPM adds in order to construct Ising ground states seems to have no such monotonicity property in its weight increments, and indeed some APs decrease the weight of the matching. We do not know of a ``physical'' interpretation of the way the Blossom algorithm chooses its APs (for any class of graphs).

\subsection{Optimized defects and backwards RG}

\begin{figure}
 \centering
 \includegraphics[width=.7\columnwidth]{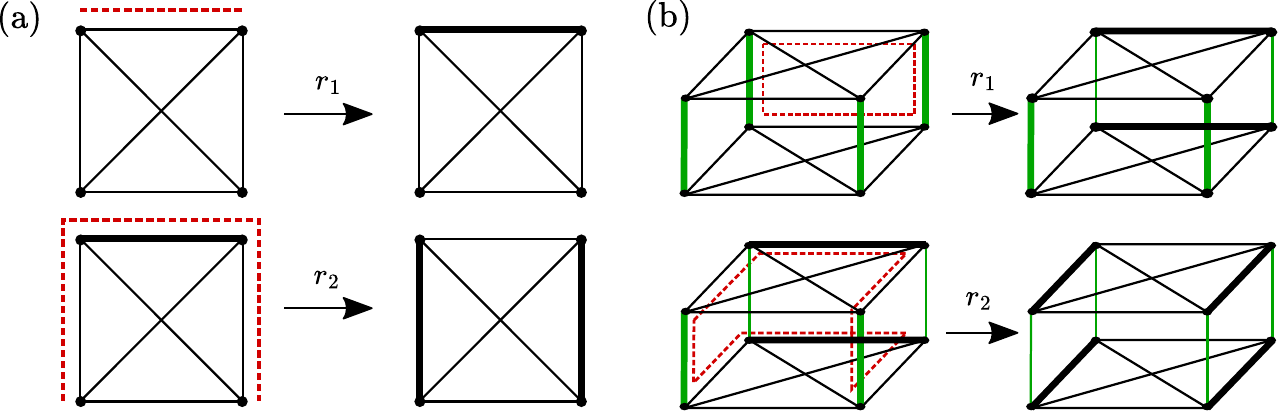}
 \caption{(a) The RG procedure on a graph $\GG_T$ of four frustrated plaquettes. Red dotted lines show augmenting paths that are formed during the RG. Bold edges indicate matchings at each step. (b) The duplicated Toulouse graph $\GG_T^2$ for the same configuration of frustrated plaquettes. Edges in green have weight $W$, which is increased gradually. Red dotted lines show cycles which are flipped, which correspond to augmenting paths in each of the copies of $\GG_T$. Bold edges indicate membership in the partial matching, which becomes complete by the end of the procedure.} 
 \label{fig:greedy_matching}
\end{figure}

Using the language of Toulouse matchings, we can define the optimized defect energy as
\begin{equation}\label{eq:optimized_defect_definition}
 \rmax = \max_{p,q } \left( E(\GG_T) -E(\GG_{T, \setminus \{p,q\}}) \right). 
\end{equation}
where $\GG_{T,\setminus \{p,q\}} $ is a Toulouse graph with plaquettes $\mathcal{P}\setminus \{p,q\} $. We will now show that $\rmax$ thus defined is indeed equal to the final increment of the RG $r_{N_f/2}$.

It will be useful to describe a concrete implementation for carrying out the procedure. Let $\GG_T$ be the Toulouse graph for a given disorder realization. We construct a ``duplicated Toulouse graph'' $\GG_T^2$, illustrated in Fig.~\ref{fig:greedy_matching}b, consisting of two identical copies of $\GG_T$, with the two copies of each vertex joined by an edge of weight $W$. 

When $W=0$, the MWPM on $\GG_T^2$ is just the set of inter-copy edges, and has weight $0$. As $W$ is gradually increased, the MWPM on $\GG_T^2$ will change whenever it becomes energetically favorable to flip matchings along a cycle in $\GG_T^2$ that traces out two copies of an AP in the two copies of $\GG$, along with two of the inter-copy edges. In fact, the value of $W$ when such a cycle is flipped is precisely the $r_n$ corresponding to the AP seen in a single copy of $\GG_T$. For large enough $W$ the MWPM in $\GG_T^2$ will consist of two copies of the MWPM in each of the copies of $\GG_T$, with none of the weight-$W$ edges matched. Sweeping through $W$ from small to large makes a sequence of minimally augmenting paths through each copy of $\GG_T$. Given the monotonicity of the sequence of $r$'s, as shown above, the values of $W$ at which the MWPM of $\GG_T^2$ changes constitute the increasing sequence of $r$'s. 


If we start with large $W$, such that the matching in each copy of $\GG_T$ corresponds to the ground state of $\mathcal{H}$, then the largest $W$ at which the MWPM on $\GG_T^2$ changes is the $W$ at which the final increment was made in the matching of $\GG_T$. By construction, since we are at the largest $W$ at which the MWPM on $\GG_T^2$ changes, the change in matchings in each of copy of $\GG_T$ amounts to inserting a pair of unpaired monomers that decreases the ground-state energy by the maximum possible amount, in both copies of $\GG_T$. Therefore the last $r$ found in the RG is the optimized defect $\rmax$ as defined in Eq.~\eqref{eq:optimized_defect_definition}. As $W$ is decreased from $\infty$ to $0$, we run the RG in reverse. This action has a natural interpretation in terms of optimized defects: each ``de-augmentation'' inserts an optimized defect in a (partial) matching.

\subsection{Optimized defects from Kasteleyn matchings}
How are we to numerically evaluate $\rmax$? Any procedure involving Toulouse graphs, especially two copies thereof, is slow (runtime scaling as a large power of $L$). We now show that $\rmax$ can in fact be efficiently calculated via a decoration of the Kasteleyn MWPM of Appendix \ref{app:kasteleyn_matching}. 

In particular, we show that (1) the largest possible decrease in energy of a fixed initially-ground-state spin configuration under flipping the signs of a string of couplings between $p$ and $q$ on the dual lattice is equal to (2) the change in ground state energy from flipping the signs of couplings along {\it any} path between $p$ and $q$ and allowing the ground state to concomitantly change. We will then show that the MWPM algorithm can be made to evaluate (1), and it therefore evaluates optimized defect energies as defined in terms of (2): optimal unfrustrations of the ground state.

In order to explain the graph construction, we first discuss 
the change in Ising ground state that results from eliminating frustrated plaquettes $p$ and $q$ in $\mathcal{H}$ by flipping the signs of the $J_{ij}$ across some path $\Pi$ from $p$ to $q$ on the dual square lattice. Call this new set of couplings $J^\Pi$.


Let us define $\Delta[\sigma, J, \Pi] \equiv -\sum_{ij\in \Pi} J_{ij} \sigma_i \sigma_j $ such that the energy change of configuration $\sigma$ from flipping the bonds that cross $\Pi$ is
\begin{equation}\label{eq:sameS}
 E_J[\sigma] - E_{J^\Pi}[\sigma] = 2\Delta[\sigma, J, \Pi].
\end{equation}
Say the ground-state pair of spin configurations for couplings $J$ is $\pm \sigma^g$, 
and consider a state $\sigma^{g'}$ that differs from $\sigma^g$ by the spins enclosed by the set of domain walls made by the symmetric difference
of $\Pi$ with another path from $p$ to $q$, which we call $\Pi'$. 
Then $\sigma^{g'}_{i} \sigma^{g'}_{j}$ differs from $\sigma^g_i \sigma^g_j$ only across $(\Pi\cup \Pi')\setminus (\Pi\cap \Pi')$, whereas $J^\Pi$ differs from $J$ across $\Pi$. Thus
\begin{align}
 E_{J^\Pi}[\sigma^{g'}] &= -\sum_{ij \notin \Pi\cup \Pi'} J_{ij} \sigma^g_i\sigma^g_j - \sum_{ij \in \Pi \setminus (\Pi\cap \Pi')} (-J_{ij})(- \sigma^g_i \sigma^g_j) -\sum_{ij \in \Pi' \setminus (\Pi\cap \Pi')} J_{ij} (- \sigma^g_i \sigma^g_j) - \sum_{ij \in \Pi \cap \Pi'} (-J_{ij}) \sigma^g_i \sigma^g_j\\
 &= E_J[\sigma^g] + 2 \sum_{ij \in \Pi'} J_{ij} \sigma^g_i \sigma^g_j \\
 &= E_J[\sigma^g] - 2 \Delta[\sigma^g, J, \Pi'] \label{eq:gsJP1} 
\end{align}
It follows that, if 
\begin{equation}\label{eq:pstar}
 \Pi^* = \text{argmax}_{\Pi'} \Delta[\sigma^g, J, \Pi'],
\end{equation}
overturning the spins enclosed by $(\Pi\cup \Pi^*)\setminus(\Pi\cap \Pi^*)$ gives the ground state of $J^\Pi$. 
We also see from \eqref{eq:gsJP1} that $\Pi^*$ is the path across which flipping the signs of the $J$ maximizes the energy decrease of $\sigma^g$. Therefore we have shown that both (1) the maximal energy decrease of $\sigma^g$ under flipping a path between $p$ and $q$ and (2) the change in ground state energy from flipping the bonds across $\Pi$---for any $\Pi$ that connects $p$ and $q$---are equal to $2\Delta[\sigma^g,J,\Pi^*]$.

Here we have discussed paths between a single pair of frustrated plaquettes, but to calculate the optimized defect, we should further optimize $\Pi$ over choices of all possible pairs $p,q \in \mathcal P$ that it connects. 

This brings us to the matching construction: starting with the dual Kasteleyn graph of Appendix \ref{app:kasteleyn_matching}, consider a graph which contains two extra vertices $s$ and $t$, each connected with zero edge weights to all $4 N_f$ vertices of the original graph that belong to Kasteleyn cities associated with frustrated plaquettes, and solve for MWPM on this graph. It is clear that the two perfect matchings, respectively $\MM$ and $\MM'$, differ (in addition to intra-city rearrangements of zero-weight bonds) by a path that comprises one bond each incident on $s$ and $t$, connections to cities for the respective pair of plaquettes $p$ and $q$, and a path on the dual Kasteleyn lattice that has intersection $\Pi$ with the dual Ising lattice. Moreover, any $\Pi$ can be lifted to a valid path on the dual Kasteleyn lattice by inserting the appropriate intra-city edges. The problem of finding the new MWPM is reduced to finding the $\Pi^*$ along which a matched-unmatched flip has the most negative cost---the same negative energy cost that would be incurred from flipping the signs of the Ising bonds across $\Pi^*$ while maintaining the ground state spin configuration induced by $J$---and therefore also the change in ground state energy from flipping the bonds across any $\Pi$ between $p$ and $q$.

Summarizing: to calculate $\rmax$ for a given set of couplings, we have to solve two MWPM problems: one being the usual Ising ground-state problem on the dual Kasteleyn lattice, and one on a graph where all frustrated cities are connected by zero-weight edges to two auxiliary vertices; the difference of MWPM energies gives us $\rmax$. This algorithm was used for all our numerics on optimized defects.

Since the just-discussed construction also tells us which plaquettes should be unfrustrated for the optimized defect, performing said unfrustration and finding the new optimized defect would provide information on the penultimate step of the RG, and so on: we could iterate to construct the sequence of $r$'s and associated APs for the whole RG.

It is worth noting that there is also a duplicated Kasteleyn version of the duplicated Toulouse graph we discussed above, in which one could imagine efficiently ``stepping through'' many steps of the RG by incrementing the inter-copy weight by a suitable amount. If the steps are fine enough that the APs created by them are spatially separated, this would be an efficient way of extracting information on the APs produced in many intermediate steps of the RG at once.

Because all our optimized defect numerics are done with fully periodic boundary conditions, both the ground state and the placement of the defect are also optimized over boundary conditions (by optimizing over which way the defect path is ``wrapped'' around the torus); see the discussion of extended ground states above.

\subsection{Optimized defects and spectra}

Here we show, adapting an argument by Motrunich, Damle and Huse who studied bipartite matchings~\cite{MotrunichDamleHuse}, that for any graph $\GG=(\VV,\EE)$ which admits a perfect matching, the optimized defect energy, defined as
\begin{equation}\label{eq:def_opt_def_for_MDH}
 \varrho = \max_{v,w \in \VV} \left[E(\GG) - E(\GG_{\setminus\{ v,w\}) }\right],
\end{equation}
(where $\GG_{\setminus\{ v,w\}}$ is the subgraph of $\GG$ induced by $\VV\setminus\{ v,w\}$) is related to the smallest-magnitude eigenvalue of any antisymmetric matrix of the form
\begin{equation}
 \Gamma_{vw} = e^{-\beta \zeta_{vw}}, \quad \Gamma_{wv} = -\Gamma_{vw},
\end{equation}
 as $ \min |\spec \bf{\Gamma}| \sim $ $e^{-\beta \varrho}$ in the $\beta\to \infty$ limit, where we ignore prefactors multiplying exponentials. More precisely, if $\gamma$ is the smallest-magnitude eigenvalue,
 \begin{equation}\label{eq:minspecgammathm}
 \varrho = \lim_{\beta \to \infty} \frac{\log (|\gamma|^{-1})}{\beta}.
 \end{equation}
 (This would be another way of showing the correspondence between $\emin$ and $\rmax$ in the Ising case, using $\bf \Gamma = \bf \Upsilon$.)
 
The Pfaffian is defined as~\cite{McCoyWuBook}
\begin{equation}
 \Pf {\bf \Gamma} = {\sum_{\MM}} (-1)^{\MM} \Gamma_{\MM(1) \MM(2)}\cdots \Gamma_{\MM({|V|-1}) \MM(|V|)},
\end{equation}
where the sum is over unique matchings $\MM$ of $\GG$, regarded as permutations of $\VV$, so that $(-1)^\MM$ is the sign of the permutation. 

When $\beta \to \infty$ the sum is dominated by the MWPM, up to a sign: $|\Pf {\bf \Gamma}| \sim e^{-\beta E(\GG)}.$ 
Observe that, even though the graph $\GG$ does not generically admit a Pfaffian orientation---if it did, we could evaluate the dimer partition function on it via a Pfaffian for arbitrary temperatures~\cite{McCoyWuBook}---the zero-temperature limit of the dimer partition function on $\GG$, thus the MWPM energy, can nevertheless be obtained using a Pfaffian. 

Next, we observe that the elements of the inverse of the matrix are related to defects: 
\begin{equation}\label{eq:UpsilonInverse}
 |({\bf \Gamma}^{-1})_{vw}| = \left| \frac{\Pf {\bf \Gamma}_{\setminus vw}}{\Pf {\bf \Gamma}} \right| \sim e^{ \beta(E(\GG) - E(\GG_{\setminus \{v,w\}})}
\end{equation}
where ${\bf \Gamma}_{\setminus vw}$ is the matrix ${\bf \Gamma}$ with the $v$th and $w$th rows and columns deleted. 

The smallest-magnitude eigenvalue of ${\bf \Gamma}$ is given by the matrix norm (say the 2-norm) of ${\bf \Gamma}^{-1}$: $ \min |\spec {\bf \Gamma}| = \gamma \sim 1/\|{\bf \Gamma}^{-1}\|$. In the $\beta\to \infty$ limit, we further have $\|{\bf \Gamma}^{-1}\| \sim \max_{vw} |({\bf \Gamma}^{-1})_{vw}| $. So, comparing Eqs.~\eqref{eq:def_opt_def_for_MDH} and \eqref{eq:UpsilonInverse}, we arrive at the desired Eq.~\eqref{eq:minspecgammathm}.

\section{More on perturbation theory}\label{app:more_fermion}

Here we discuss the spectrum of the matrix $\Upsilon$ defined in the main text. This spectrum is given by $\{\pm e^{-2\beta R_n} \}$ for $n \in \{ 1, \cdots, N_f/2\}$. The energy increments of our RG in the Ising/matching space are given by $\{2 r_n\}$. The Ising ground-state energy of $\mathcal{H}$ can be expressed in terms of either of these sequences, as we have stated, so that
\begin{equation}\label{eq:rsum_equiv_Rsum}
 \sum_{n = 1}^{N_f/2} R_n = \sum_{n = 1}^{N_f/2} r_n.
\end{equation}
Our goal now is to show that the sequences are identical, $R_n = r_n$ for all $n\in\{1,\dots,N_f/2\}$. Throughout this section, we will drop all factors that multiply exponentials of $\beta$.

Paralleling the RG construction of the sequence of Ising Hamiltonians $\{\mathcal{H}_n\}$ containing plaquettes $\{\mathcal{P}_n \}$, consider the corresponding sequence $\{\boldups_n \}$ such that the $N$th member of this sequence, the $2N \times 2N$ matrix $\boldups_N$, has elements $ \Upsilon_{N, pq} \sim e^{-2\beta\zeta_{pq}}$ between plaquettes $p, q$ in $\mathcal{P}_N$.

We will show inductively that, for all $N \in \{1, \cdots, N_f/2\} $,
\begin{enumerate}
 \item $\boldups_N$  has spectrum $\{ \pm e^{-2\beta r_n} \}$, and that
 \item In the $\beta \to \infty$ limit the $N$th pair of eigenstates is localized on $p_N$ and $q_N$: these are the two plaquettes added in this RG step.
\end{enumerate}

The base case $N = 1$ is trivial: the first RG step matches the plaquettes $p_1, q_1$ of $\mathcal{H}$ with the smallest distance between them, $r_1 = \zeta_{p_1 q_1} < \zeta_{p, q}$ for all $\{p, q\} \neq \{ p_1, q_1\}$. The corresponding matrix element $e^{-2\beta r_1}$ is then parametrically larger than all others as $\beta \to \infty$, and we are guaranteed a pair of eigenstates localized on $p_1$ and $q_1$ with eigenvalues $\pm e^{-2\beta r_1}$.

Assuming that statements 1 and 2 are correct for $\boldups_{1}, \cdots, \boldups_{N-1}$, we have to deal with
\begin{equation}
\scalebox{1.5}{$\boldups_{N}$} = 
\left(
\begin{array}{cccccccccc|cc}
 & & & & & & & & & & e^{-2\beta \zeta_{p_1 p_N}} & e^{-2\beta \zeta_{p_1 q_N}} \\
 & & & & & & & & & & e^{-2\beta \zeta_{q_1 p_N}} & e^{-2\beta \zeta_{q_1 q_N}} \\
 & & & & & & & & & & \vdots & \vdots \\
 & & & & & \scalebox{1.5}{$\boldups_{N-1}$} & & & & & e^{-2\beta \zeta_{p_n p_N}} & e^{-2\beta \zeta_{p_n q_N}} \\
 & & & & & & & & & & e^{-2\beta \zeta_{q_n p_N}} & e^{-2\beta \zeta_{q_n q_N}} \\
 & & & & & & & & & & \vdots & \vdots \\
 \hline
 & & & & & & & & & & 
0 & e^{-2\beta \zeta_{p_N q_N}} \\
 & & & & & \mathrm{h.c.} & & & & & \rm{h.c.} & 0 \\
\end{array}
\right).
\end{equation}
This matrix is in the basis ordered as $p_1, q_1, \cdots, p_N, q_N$.
The newly added elements come from the distance between the new plaquettes $p_N, q_N$, as well as the distances between these and the plaquettes already in $\mathcal{P}_{N-1}$.

The optimization over augmenting paths embedded in the RG puts constraints on these new matrix elements. Consider the RG step $n < N$ at which plaquettes $p_n, q_n$ were included in the matching. At this point a candidate RG step would have been to directly match $p_n$ to $p_N$, since $p_N$ was unmatched then. The augmenting path that was chosen between $p_n$ and $q_n$ must have produced a smaller increment than this direct matching. 
Therefore $r_n < \zeta_{p_n p_N}$. Similarly $r_n < \zeta_{p_n q_N}, \zeta_{q_n p_N}, \zeta_{q_n q_N}$, and because $p_N, q_N$ were not matched to each other until this RG step, we have $\zeta_{p_N q_N} > r_{N-1} $ (and hence greater than all previous $r_n$, due to the monotonicity $r_n < r_{n+1}$).
 
We assumed for each $n \leq N-1$ that, as $\beta \to \infty$, the energies $\pm e^{-\beta r_n}$ are associated with wavefunctions localized on $p_n, q_n$. From the argument of the previous paragraph, we see that the matrix elements between these plaquettes and $p_N, q_N$ in $\boldups_N$ are parametrically  smaller than $e^{-2\beta r_n}$ as $\beta \to \infty$, so each of the heretofore created $e^{-2\beta r_n}$, i.e. the spectrum of $\boldups_{N-1}$, must continue to be in the spectrum of $\boldups_N$. Now note that the relation \eqref{eq:rsum_equiv_Rsum} holds for any upper limit of the sum between $1 $ and $N_f/2$. Knowing that $R_n = r_n$ for all $n \leq N-1$, we can deduce from \eqref{eq:rsum_equiv_Rsum} that $R_N = r_N$. Moreover, because of the vanishing (compared to $e^{-2\beta r_n}$) coupling between each $p_n, q_n$ and $p_N, q_N$, the $n$th pair of eigenstates continues to be localized on the former pair of plaquettes, while the new eigenstates will be localized on $p_N, q_N$ (with support between them predominantly on the alternating path that the RG chooses at step $N$).

\section{One dimension}\label{app:1d}

The simple problem of a random tight-binding chain~\cite{Dyson}, 
\begin{equation}
 {\HH}_{1d} = \sum_i e^{-2\beta \zeta_i}( \ket{i}\bra{i+1} + h.c.)
\end{equation}
is readily solved in the large-$\beta$ limit using our RG, which now becomes equivalent to a familiar strong-disorder RG on a chain~\cite{FisherTFIM}.

Each step of the RG finds the largest product of the form $e^{-2\beta(\zeta_i - \zeta_{i+1}+\dots -\zeta_{i+2\ell -1}+\zeta_{i+2\ell})}$, and makes a wavefunction with this energy mostly localized on sites $i$ and $i+2\ell+1$ . This procedure corresponds to a rather involved way of solving the matching problem. (An even-site periodic chain admits just two matchings!) 

The chain ${\HH}_{1d}$ can be seen as the effective problem living on the frustrated plquettes of a very special 2d Ising model, with frustrated plaquettes and their matchings restricted to one row of the dual lattice, e.g. by making all bonds have $J=\infty$ other than the ones on this row, which are drawn from the usual Gaussian $P(J)$. Then $\Upsilon$ truncated to nearest neighbors (which is sufficient, since it is never energetically favorable to match further neighbors) has the form of ${\HH}_{1d}$ discussed above, with $\mu > 0$ yielding distributions of $\{\zeta_i\}$ that alternate among successive plaquette-to-plaquette distances, since frustrated plaquettes will tend to occur in co-located pairs. 

Optimized defects come from exponentially rare regions dimerized ``the wrong way'', giving $\expect[\rmax] \sim 
z \log L$, with $z\to \infty$ as the dimerization vanishes, $\mu \to 0^+$. 
With $\mu = 0$, $\{\zeta_i\}$ are i.i.d., and the last increment of the RG, $\rmax \equiv r_{N_f/2}$, as an appropriate alternating sum, is easily seen to be $\rmax \sim L^{1/2}$---the much-studied $\psi=1/2$ infinite-randomness scaling in 1d.
In our 2d problem, the Griffiths phase obtains for large $\mu$ (in the ferromagnetic phase) as in the 1d problem, but  as $\mu$ is decreased there is a transition (with diverging $z$) at positive $\mu$ to a different phase; such a transition only occurs in the 1d problem at $\mu=0$.

\end{spacing}

\end{document}